\documentclass[aps,twocolumn,showpacs]{revtex4}
\usepackage{amssymb}
\usepackage{graphicx}

\def\be{\begin{equation}}
\def\ee{\end{equation}}
\def\bea{\begin{eqnarray}}
\def\eea{\end{eqnarray}}

\newcommand{\w}[1]{(\ref{#1})}

\newcommand{\ket}[1]{\mbox{$| #1 \rangle$}}

\newcommand{\proj}[1]{\mbox{$| #1 \rangle \! \langle #1 |$}}

\begin{document}

\title{Secrecy content of two-qubit states}
\author{A. Ac\'{\i}n$^{1}$, J. Bae$^{1}$, E. Bagan$^{2}$, M. Baig$^{2}$,
Ll. Masanes$^{3,4}$, R. Mu\~{n}oz-Tapia$^{2}$}

\affiliation{ $^{1}$ICFO-Institut de Ci\`encies Fot\`oniques,
Jordi Girona 29, Edifici Nexus II, 08034 Barcelona, Spain\\
$^2$Grup de F{\'\i}sica Te{\`o}rica, Universitat
Aut{\`o}noma de Barcelona, 08193 Bellaterra (Barcelona), Spain\\
$^3$Dept. d'Estructura i Constituents de la Mat\`eria, Univ. de Barcelona, 08028 Barcelona, Spain\\
$^4$School of Mathematics, University of Bristol, Bristol BS8 1TW, United Kingdom}
\date{\today }

\begin{abstract}
We analyze the set of two-qubit states from which a secret key can
be extracted by single-copy measurements plus classical processing
of the outcomes. We introduce a key distillation protocol and give
the corresponding necessary and sufficient condition for positive
key extraction. Our results imply that the critical error rate
derived by Chau, Phys. Rev. A {\bf 66}, 060302 (2002), for a
secure key distribution using the six-state scheme is tight.
Remarkably, an optimal eavesdropping attack against this protocol
does not require any coherent quantum operation.
\end{abstract}

\maketitle

\section{Introduction}

It is known that all quantum correlations can be converted into
\emph{secret} ones, namely, into correlations that cannot be
distributed by local operations and public
communication~\cite{Acin}. However, the identification of the
precise quantum correlations (entangled states) that can be
converted into a secret key remains an open problem. This paper
focuses on this problem: we wish to determine which two-qubit
states contain \emph{distillable} secret correlations. More
precisely, we aim to characterize the set of two-qubit states from
which a secret key can be extracted by \emph{SIngle-copy
Measurements plus ClAssical Processing} (SIMCAP) protocols.

As usual, Alice and Bob are the honest parties willing to
communicate secretly and Eve is the adversary who tries to learn
the secret messages. The  scenario for key extraction that we
consider is quite similar to that for entanglement distillation.
The honest parties initially share a large number, $N$, of copies
of a known two-qubit state $\rho_{AB}$. Instead of distilling
entangled bits, or singlets, Alice and Bob's task here consists in
extracting secret bits. Pure secret and entangled bits are indeed
two different information resources that can be extracted from
quantum states. Notice that in contrast with general security
proofs of quantum key distribution (QKD) protocols, it is assumed
that Alice and Bob know they share $N$ independent copies of the
two-qubit state $\rho_{AB}$. We restrict our considerations to
SIMCAP protocols for several reasons. First, they do not require
any coherent quantum operation, so they are experimentally
feasible with present day technology. Second, it is interesting to
compare these protocols with those employing coherent quantum
operations performed by Alice and Bob. Finally, results obtained for quantum
states in the SIMCAP scenario can be applied to quantum channels
and \emph{prepare and measure} QKD schemes~\cite{p&m}, such as
BB84~\cite{BB84}.

%In order to achieve our task,
In this paper, we consider a slightly improved version of the
SIMCAP protocol with two-way communication introduced
in~\cite{AGM}. We derive a necessary and sufficient condition any
two-qubit state must satisfy for this protocol to be secure. The
sufficiency of this condition is proved by showing that, if it
holds, our protocol enables extracting a key that is secure against
any attack. The necessity follows from the existence of an
explicit eavesdropping attack (given below), that breaks the
protocol if the aforementioned condition is not satisfied.
Remarkably enough, Eve can implement this attack without any
coherent quantum operation. As far as we know, this is the first
necessary and sufficient condition for key distillation from
quantum states
using a two-way communication SIMCAP protocol. %In addition, this is the
%quantum-cryptographic scheme ---in the SIMCAP scenario--- more
%robust in terms of noise.

%Chau presented in~\cite{Chau} a security proof (without
%assumptions) for the six-state protocol with two-way
%communication. He proved the security of his scheme for error rates (QBER) smaller than
%27\%. The necessary part of condition can be used to show
%that this error rate is the maximal attainable with Chau's protocol.

\section{The protocol}

As mentioned above, Alice and Bob share $N$ copies of a known
bipartite state $\rho_{AB}$. Given this assumption, the SIMCAP key
distillation protocol consists of three steps:
%  \begin{enumerate}
 {(i)}~local measurement on each qubit pair $\rho_{AB}$,
 {(ii)}~advantage distillation, and
 {(iii)}~one-way key extraction.
 % \end{enumerate}

{\sl Measurements:} Step {(i)} can be decomposed into the
operations {(a), (b), and (c)} defined as follows.
%  \begin{enumerate}
%
Operation {(a)}~is a single-copy filtering operation that Alice
and Bob perform in order to
  maximize the entanglement of formation of their state. The operators $F_A$ and
  $F_B$ that characterize this filtering,
  $\rho_{AB}\to F_A \otimes F_B\ \rho_{AB}\ F_A^\dagger \otimes F_B^\dagger$,
  are described in~\cite{V}. If the filtering fails, the qubit pair
  is rejected. If it succeeds, the resulting two-qubit state is
  diagonal in the Bell basis (defined in~\cite{bb}):
    \be
   \rho_{AB}\to \lambda_1 \left[\Phi^+\right] + \lambda_2 \left[\Phi^-\right] +
    \lambda_3 \left[\Psi^+\right] + \lambda_4 \left[\Psi^-\right],
    \label{bell}
    \ee
  with $\lambda_i \geq 0$ and $\sum_i \lambda_i=1$. Throughout this paper square brackets  denote one-dimensional
  projectors (not necessarily normalized); e.g.,
  $[\psi]=\proj{\psi}$.
  If~$\rho_{AB}$ is already diagonal in the Bell basis, this
  filtering leaves it unchanged.
 Operation {(b)} is a local unitary transformation that Alice and Bob  apply to the state \w{bell}
 to ensure that
    \be
    \lambda_1= \max_i \lambda_i ,\qquad
    \lambda_2= \min_i \lambda_i \ . \label{minmax}
    \ee
 This is just a permutation of the coefficients $\lambda_i$   in \w{bell},
 and {\em any} such permutation
 can  be achieved using only local unitaries~\cite{permutations}.
  One can thus associate to any two-qubit state $\rho_{AB}$ the pair
  of coefficients~$(\lambda_1 , \lambda_2)$.
The last operation, (c), consists in measuring each qubit in the computational
basis~$\{ \ket{0},\ket{1} \}$.

%  \end{enumerate}
These three operations can be seen as a single measurement
performed by each of the honest parties, with outcomes: \verb"0",
\verb"1" and \verb"reject". After discarding all the instances
where the outcome \verb"reject" is obtained, each of the honest
parties has  a list of partially correlated bits. These two lists
do not constitute a shared secret key yet, because in general they
are neither equal nor secret. Our goal is now to distill them to a
secret key [steps (ii) and (iii) above].

{\sl Advantage distillation:} Step (ii) is a reconciliation scheme
introduced by Maurer in \cite{Maurer} that uses two-way
communication. Within this scheme, each of the honest parties
transforms blocks of $M$ bits into a single bit. By doing that,
Alice and Bob map their initial lists of bits into shorter, more
secret and correlated ones. To achieve this goal, Alice randomly
chooses $M$ bits from her list of accepted outcomes, and Bob takes
their $M$ counterparts from his list:
    \bea
    A_1,A_2 \ldots A_M \nonumber\\
    B_1,B_2 \ldots B_M.
    \label{lists}
    \eea
%In our problem, each pair $(A_i,B_i)$ is independent of the rest.
Next,
Alice generates a secret random bit $s_A$, computes the $M$
numbers $X_i := (A_i+s_A) \bmod 2$, and sends
the $M$-bit string
  \be
  X_1,X_2, \ldots, X_M
  \label{publicmessage}
  \ee
through the insecure but authenticated public channel. Bob then
adds bitwise (mod~2) this string to his list, $B_1,B_2,\ldots
B_M$. If he obtains the same result $s_B$ for the $M$ sums, i.e.,
if $(B_i+X_i) \bmod 2=s_B$ for $i=1,2,\ldots M$, he keeps the bit
$s_B$ and communicates its acceptance to Alice. Otherwise, the two
parties reject the $M$ bits.
The bits $s_A$ and $s_B$ are the result of the advantage
distillation process~(ii). A large number of pairs $(s_A,s_B)$ constitute
the input of~(iii).

{\sl One-way key extraction:} Step~(iii) consists of the one-way
communication procedure given by Devetak and Winter
in~\cite{Winter}. It concerns the situation where Alice has a
classical random variable correlated to Bob and Eve's quantum
states, and it enables (when possible) transforming these
classical-quantum-quantum (CQQ) correlations into a secret key
with maximal rate. In our case, the honest parties have the
classical random variables $(s_A,s_B)$ correlated with Eve's
quantum states (CCQ correlations), this being a particular case of
the scenario considered in~\cite{Winter}. Thus, their techniques
immediately apply.

Having discussed the protocol with some detail, we now state our
main result: {\em A secret key can be extracted from a two-qubit
state $\rho_{AB}$ by the  protocol above if and only if its
associated weights $(\lambda_1,\lambda_2)$ satisfy}
  \be
  (\lambda_1 - \lambda_2 )^2>
  (1-\lambda_1-\lambda_2)(\lambda_1+\lambda_2)
  \label{condition}
  \ee
The sufficient and necessary statements of this result are proved in sections III and IV,
respectively.

\section{Security proof}

Let us first prove the security of this protocol. As usual, we
conservatively assume that Eve has a large quantum system
that is a purification of the whole state $\rho_{AB}^{\otimes N}$. Note that all
purifications of Alice and Bob's state are equivalent, since
they only differ by a local unitary operation on Eve's Hilbert space.
Without any loss of generality, the state of the three parties can be taken  to be
$\ket{\Psi_{ABE}}^{\otimes N}$, where $\ket{\Psi_{ABE}}$ is a purification
of $\rho_{AB}$, i.e.,
$
  \rho_{AB}=\mbox{tr}_E [\Psi_{ABE}]
$.
After the filtering operation~(a), the tripartite state
is still pure, hence, Eve holds the system that purifies the
Bell-diagonal state \w{bell}. The three parties thus share
many copies of the state
  \bea
&& \kern-1.5em F_A\!\! \otimes\!F_B\!\otimes \!I\!_E\ket{\Psi\!_{ABE}} % \nonumber\\
 \!=\! \sqrt{\lambda_1}\ket{\Phi^+}\!_{AB}\ket{1}\!_E\!+\!
  \sqrt{\lambda_2}\ket{\Phi^-}\!_{AB} \ket{2}\!_E\nonumber\\
&&\kern-1.5em+
 \sqrt{\lambda_3}\ket{\Psi^+}\!_{AB}\ket{3}\!_E \!+\!
  \sqrt{\lambda_4}\ket{\Psi^-}\!_{AB}\ket{4}\!_E   .
  \eea
After step~(i), Alice and Bob are left with classical data,
whereas Eve could still hold a quantum system. The correlations
they share are described by the state (up to normalization)
  \be
  \sum_{x}\ [x]_{AB}\otimes[\psi_x]_E,
  \nonumber
  \ee
where $x=00,01,10,11$, and
  \bea
  &&\ket{\psi_{00/11}}=\sqrt{\lambda_1}\ket{1}\pm\sqrt{\lambda_2}\ket{2} ,
  \nonumber\\
  &&\ket{\psi_{01/10}}=\sqrt{\lambda_3}\ket{3}\pm\sqrt{\lambda_4}\ket{4} .
 % \nonumber
  \eea
Notice that the above vectors are non-normalized.
%  \bea
%  &&\frac{1}{2}\, \left[00\right]_{AB} \otimes \left[\sqrt{\lambda_1}\ket{1} + \sqrt{\lambda_2}\ket{2}
%  \right]_E+ \nonumber \\
%  &&\frac{1}{2}\, \left[11\right]_{AB} \otimes \left[\sqrt{\lambda_1}\ket{1} - \sqrt{\lambda_2}\ket{2}
%  \right]_E+ \nonumber \\
%  &&\frac{1}{2}\, \left[01\right]_{AB} \otimes \left[ \sqrt{\lambda_3}\ket{3} + \sqrt{\lambda_4}\ket{4}
%  \right]_E+ \nonumber \\
%  &&\frac{1}{2}\, \left[10\right]_{AB} \otimes \left[ \sqrt{\lambda_3}\ket{3} - \sqrt{\lambda_4}\ket{4}
%  \right]_E .
%  %\label{tripartite0}
%  \eea
After step~(ii), Eve has her $M$ (four-dimensional) quantum systems as well as the
information that the honest parties have exchanged through the public
channel.
In particular, Eve has the $M$-bit string \w{publicmessage}. If she performs
the unitary transformation
  \be
  U_i=[1]_E+(-1)^{X_i}[2]_E+[3]_E+(-1)^{X_i}[4]_E
  \label{transform}
  \ee
to her $i$-th system ($i=1\ldots M$), up to normalization the
tripartite state becomes
%  \bea
%  &&\left[00\right]_{AB} \otimes \left[\sqrt{\lambda_1}\ket{1} + \sqrt{\lambda_2}\ket{2}
%  \right]_E^{\otimes M} + \nonumber \\
%  &&\left[11\right]_{AB} \otimes \left[\sqrt{\lambda_1}\ket{1} - \sqrt{\lambda_2}\ket{2}
%  \right]_E^{\otimes M} + \nonumber \\
%  &&\left[01\right]_{AB} \otimes \left[ \sqrt{\lambda_3}\ket{3} + \sqrt{\lambda_4}\ket{4}
%  \right]_E^{\otimes M} + \nonumber \\
%  &&\left[10\right]_{AB} \otimes \left[ \sqrt{\lambda_3}\ket{3} - \sqrt{\lambda_4}\ket{4}
%  \right]_E^{\otimes M} .
%  \label{tripartite}
%  \eea
  \be
  \sum_x[x]_{AB}\otimes [\psi_x]_E^{\otimes M} .
  \label{tripartite}
  \ee
%The transformation \w{transform} does not change Eve's status
%since it is reversible. However,
 After this transformation, the
tripartite state \w{tripartite} becomes completely uncorrelated
to~\w{publicmessage}. The rest of the protocol is also independent
of~\w{publicmessage}, and this information is no longer useful.
Hence, all the correlations among Alice, Bob and Eve before
step~(iii) are described by the state \w{tripartite}.

It was proven in~\cite{Winter}, that the secret key
rate one can achieve with one-way communication ($K_{\rightarrow}$) when Alice
holds a classical system satisfies:
  \be
  K_{\rightarrow} \geq I(A:B)-I(A:E),
  \label{ratel}
  \ee
where $I(X:Y)$ is the mutual information  referred to the state
\w{tripartite}, and is defined in~\cite{mutinf}. After some
algebra, the following equality can be obtained
  \bea
 && I(A:B)-I(A:E)= 1 -h(\epsilon_B) \nonumber\\
 &&- (1-\epsilon_B)\, h\!\left( \frac{1-\Lambda_{\rm eq}^M}{2} \right)
  -\epsilon_B    \, h\!\left( \frac{1-\Lambda_{\rm dif}^M}{2} \right) ,
  \label{calcul}
  \eea
where
  \bea
 &\displaystyle \epsilon_B =\frac{(\lambda_3+\lambda_4)^M}
  {(\lambda_1+\lambda_2)^M+(\lambda_3+\lambda_4)^M} \ , &\nonumber\\
 &\displaystyle \Lambda_{\rm eq} = \frac{|\lambda_1-\lambda_2|}{\lambda_1+\lambda_2}, \quad
 \Lambda_{\rm dif} = \frac{|\lambda_3-\lambda_4|}{\lambda_3+\lambda_4}\ ,&\label{errbob}
  \eea
$h(x)=-x \log_2 x - (1-x)\log_2 (1-x)$, and the subscript `eq' (`dif') refers to the outcome $A$ being equal to (different from)~$B$. It can be checked that
if condition \w{condition} is satisfied, there exists a
sufficiently large $M$ such that  the right-hand side of
\w{ratel}, i.e., Eq.~(\ref{calcul}), is positive. Thus, a secret
key can be extracted from $\rho_{AB}$ with our SIMCAP
protocol. This completes the security proof. In the next section
we prove that condition \w{condition} is tight.

\section{Optimal eavesdropping attack}

Let us present a particular eavesdropping attack that is optimal
in the sense that it breaks our SIMCAP protocol
if~(\ref{condition}) is not satisfied. This attack is similar to
that in~\cite{sing}.

Without loss of generality, we assume that in step~(iii) the public
communication is sent from Alice to Bob. In the attack,
Eve makes a guess, $s_E$, for Alice's outcome $s_A$
in such a way that $s_E$ and $s_B$ are independent when
conditioned on $s_A$. That is, the probability distribution for
these random variables $P({s_A,s_B,s_E})$ satisfies
  \be
  P({s_B, s_E|s_A})=P({s_B|s_A})\, P({s_E|s_A}).
  \label{independent}
  \ee
To accomplish this, she first waits until step~(ii) is completed [recall that at this
stage the three parties share the state \w{tripartite}], and
performs the two-outcome measurement defined by the projectors
  \be
  F_{\rm eq}=[1]_E+[2]_E ,\quad%\quad\mbox{and}\quad
  F_{\rm dif}=[3]_E+[4]_E ,
  \label{projectors}
  \ee
on each one of her $M$ systems. According to \w{tripartite}, all
$M$ measurements give the same outcome.
If Eve obtains the outcome corresponding to $F_{\rm eq}$, the
tripartite state becomes (up to normalization)
 \be
 [00]_{AB}\otimes [\psi_{00}]_E^{\otimes M}
 +[11]_{AB}\otimes [\psi_{11}]_E^{\otimes M}
  \label{AeqB}
 \ee
%  \bea
%  &&\left[00\right]_{AB}\otimes\left[\sqrt{\lambda_1}\ket{1}+\sqrt{\lambda_2}\ket{2}\right]_E ^{\otimes M} +
%  \nonumber\\
%  &&\left[11\right]_{AB}\otimes\left[\sqrt{\lambda_1}\ket{1}-\sqrt{\lambda_2}\ket{2}\right]_E ^{\otimes M} .
%  \label{AeqB}
%  \eea
In order to learn $s_A$, she must discriminate between the two
pure states $\psi_{00}$ and~$\psi_{11}$. It was proved
in~\cite{error} that the minimum error probability she can achieve
is
  \be
 P^{ \rm error}= \frac{1}{2}-\frac{1}{2}
  \sqrt{1-c^{2M}} ,
  \label{e}
  \ee
where $c$ is the overlap between the states. Applying this formula
to~\w{AeqB}, we obtain the error probability in guessing $s_A$
  \be
  \epsilon_{\rm eq}=\frac{1}{2}-\frac{1}{2} \sqrt{1-\Lambda_{\rm eq}^{2M}} .
 \label{PeA=B}
  \ee
Similarly, if Eve obtains instead the outcome corresponding to
$F_{\rm dif}$,
%the tripartite state becomes, up to normalization,
%  \bea
%  &&\left[01\right]_{AB}\otimes\left[\sqrt{\lambda_3}\ket{3}+\sqrt{\lambda_4}\ket{4}\right]_E ^{\otimes M} +
%  \nonumber\\
%  &&\left[10\right]_{AB}\otimes\left[\sqrt{\lambda_3}\ket{3}-\sqrt{\lambda_4}\ket{4}\right]_E ^{\otimes M} .
%  \eea
%As before, Eve aims to learn $s_B$, that she knows to satisfy
%$s_B=1-s_A$. Applying formula \w{e} again, the error probability
%is in this case,
the error probability  $\epsilon_{\rm dif}$ is given by~(\ref{PeA=B}) with
the substitution $\Lambda_{\rm eq}\to\Lambda_{\rm dif}$.
%  \be
%  P(E\neq B|A\neq B)=\frac{1}{2}-\frac{1}{2} \sqrt{1-\Lambda_2^{2M}} .
%  \ee
%After this measurement, Alice, Bob and Eve share a classical
%probability distribution $P(A,B,E)$, where the random variable $E$
%stands for Eve's guess on $s_A$ and the outcome of the
%measurement~\w{projectors}.
%Note that in this attack, Eve measures right after the advantage
%distillation, step~(ii), although in principle, and in order to
%optimize her measurement, she could have waited until after the
%whole reconciliation process had been completed.
%We next prove that no secret bits can be extracted from $P(A,B,E)$
%if~(\ref{condition}) does not hold. To do that, we will show that
%Eve can always transform $P(A,B,E)$~into a non-distillable
%probability distribution $Q(A,B,E)$. This automatically implies
%that $P(A,B,E)$ is not distillable either.
%Eve's local map $P\rightarrow Q$ works as follows.

At this point, Eve's information consists of $s_E$ (her guess for
$s_A$) as well as the outcome of the measurement~\w{projectors}.
To ensure~\w{independent}, Eve proceeds as follows. From \w{minmax}, it can be seen that
$\Lambda_{\rm dif} \leq \Lambda_{\rm eq}$, which implies that $\epsilon_{\rm dif} \leq
\epsilon_{\rm eq}$. Then, when she obtains the outcome corresponding to~$F_{\rm dif}$, she
increases her error until $\epsilon_{\rm dif} = \epsilon_{\rm eq}$. She achieves this
by changing the value of $s_E$ with some probability. After this
operation the tripartite probability distribution is of the
form~\w{independent}. Additionally we know that $P({s_B|s_A})$
 and $P({s_E|s_A})$ are binary symmetric channels with error
 probability $\epsilon_B$ in~\w{errbob} and $\epsilon_{\rm eq}$
 in~\w{PeA=B},
 respectively. It is proven in \cite{Maurer} that in such
 situation the one-way key rate is
  \be
  K_{\rightarrow}= h(\epsilon_{\rm eq})-h(\epsilon_B),
  \label{ratel2}
  \ee
which is non-positive if
  \be
  \epsilon_{\rm eq} \leq \epsilon_B \ .
  \label{be}
  \ee
Let us finally prove that this inequality holds for all values of
$M$ if condition \w{condition} is not satisfied. Define
$z=\lambda_1+\lambda_2$. The range of interest is $1/2\leq z\leq
1$, since no secret key can be extracted from a separable
state~\cite{sep} and a two-qubit state is entangled {\em iff}
$\lambda_1>1/2$. After some algebra, one can prove the inequality
  \be
  \frac{1}{2}-\frac{1}{2}\sqrt{1-\left( \frac{1-z}{z} \right)^M} \leq
  \frac{(1-z)^M}{z^M+(1-z)^M} ,\label{prev}
  \ee
where $M$ is {\em any} positive integer. The right-hand side
of~\w{prev} is equal to $\epsilon_B$, whereas the left-hand side
is an upper bound for $\epsilon_{\rm eq}$. This bound follows from the
inequality $(\lambda_1-\lambda_2)^2/z^2 \leq (1-z)/z$, which is
the negation of~\w{condition}. In summary, if
condition~\w{condition} is not satisfied, no secret key can be
distilled with the considered protocol. Since we have previously
proven the sufficiency of~\w{condition}, the attack we have
considered is optimal and the security bound~\w{condition}  is
tight for our SIMCAP protocol.

It is worth analyzing the resources that this optimal
eavesdropping attack requires. First of all, we note that Eve
does not need to perform any coherent operation, i.e., she can make do with
single-copy measurements. This follows from the fact that
%, when distinguishing $M$
%copies of two pure states,
%consisting of the tensor-product of many copies of the same state,
%  \bea
%  &&\ket{\Gamma}=\ket{\gamma}^{\otimes M} \nonumber\\
%  &&\ket{\Delta}=\ket{\delta}^{\otimes M},\nonumber
%  \eea
the minimum error probability~\w{e} can be attained using an
adaptive discrimination protocol consisting of  projective
measurements on each one of the $M$ copies~\cite{Brody}.
Therefore, in order to break our key distillation protocol, what
Eve does need is the ability to store her quantum states until
after listening to
the (public) communication between the honest parties in step~(ii). %This is
%necessary to know which is the transformation \w{transform}.
That is to say, she requires a quantum memory. If Eve  can neither
perform coherent operations nor have a quantum memory, the
necessary and sufficient condition for the success of this
protocol is $\lambda_1>1/2$ ~\cite{AGM}, which is the entanglement
condition for two-qubit states.

\section{Final remarks}

In this paper, we have considered the problem of secret key extraction
from two-qubit correlations. We have derived the necessary and sufficient condition for
positive key rate using an improved version of the SIMCAP protocol
of Ref.~\cite{AGM}. If this condition does not hold, we have shown that an optimal
attack can be implemented without any coherent quantum operation.
In this case, and contrary to what happens in~\cite{qm}, quantum
memory gives a significant advantage to Eve.
\begin{figure}[h]
%  Requires \usepackage{graphicx}
 % \includegraphics[width=8 cm]{wernersec}\\
  \setlength{\unitlength}{1.9em}
\thinlines
\begin{picture}(12,5)(0,0)
%\multiput(0,0)(1,0){13}{\line(0,1){5}}
%\multiput(0,0)(0,1){6}{\line(1,0){12}}
\put (0,0){ \includegraphics[width=7.37cm]{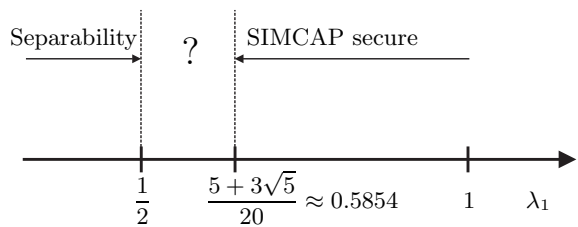}} \put
(2.55,0.6){$\displaystyle{1\over2}$}
 \put (4.1,0.6){$\displaystyle{5+3\sqrt{5}\over20}\approx 0.5854$}
 \put (9.65,0.6){$1$} \put (11,0.6){$\lambda_{1}$} \put
(5.0,4.1){SIMCAP secure} \put (3.6,3.7){{\Large ?}} \put
(-0.1,4.1){Separability}
\end{picture}\\
  \caption{Security bounds for Werner states
  of two qubits, Eq.~ (\ref{werner}). If Alice and Bob can apply coherent operations,
  all entangled states are distillable to a secret key through
  entanglement distillation. Here, we show the security of entangled
  Werner states under SIMCAP protocols up to $\lambda_1\approx 0.58$.
  Whether the gap in the figure can be closed remains an open question.}
  \label{wernersec}
\end{figure}

In view of the above, the first natural question one can ask
concerns the optimality of the SIMCAP protocol discussed here. In
other words, does condition \w{condition} characterize the set of
all distillable two-qubit states with SIMCAP protocols? Let us
argue that this could indeed be the case. Recall that our protocol
consists of three steps: measurements followed by two-way and
one-way reconciliation. Concerning the third step, we employ the
optimal protocol~\cite{Winter}. Therefore, the weak part in the
reconciliation process corresponds to the two-way communication
step. Here, we have used the standard advantage distillation
protocol. Notice that its coherent version, usually called
recurrence, combined with one-way hashing techniques, enables the
distillation of pure-state entanglement from any entangled
two-qubit state~\cite{distill}. As far as the measurement part is
concerned, the single-copy filtering operation~(a) is optimal in
terms of entanglement enhancing \cite{V}. Moreover, we have
numerically checked that for
Bell diagonal states of the form~(\ref{minmax}),
measuring in the computational basis  is optimal within
our reconciliation scheme. All this suggests that the necessary
and sufficient condition~(\ref{condition}) could very well be
completely general. If this were the case, there would exist some
two-qubit entangled states for which extracting secret bits would
require coherent operations (see Fig. \ref{wernersec}). In other
words, there would be quantum states whose secrecy content would
not be distillable by SIMCAP protocols.

%Therefore, it would be interesting to relate condition
%\w{condition} with other meaningful two-qubit state and one-qubit
%channel properties.

%Recall that from any two-qubit entangled state secret bits can be extracted
%through entanglement distillation~\cite{distill}. As said, the
%considered SIMCAP protocol succeeds for all two-qubit entangled
%states if Eve is restricted to measure before the reconciliation
%process has started~\cite{AGM}. Therefore, a natural open question
%raised by our analysis is whether quantum coherent operations are
%really required for a positive key extraction from any entangled
%two-qubit state.

%Because this protocol is implemented with SIMCAP, we have also
%found which quantum channels allow the performance of the
%equivalent {\em prepare and measure} protocol. When the generation
%of secret key with a given quantum state/channel is not possible,
%the eavesdropping attack preventing it can be implemented with
%single-copy operations and quantum memory. In the case where the
%eavesdropper is restricted to not having quantum memory, secret
%key can be generated from all quantum correlations/channels.

Since our protocol does not require any coherent quantum operation
on Alice and Bob's side, our results can be related to the
security of prepare and measure schemes, such as BB84~\cite{BB84}.
Indeed, every state can be associated to a channel, and then, the
sequence of measurements defines a QKD prepare and measure
protocol. Note however that in a fully general security proof for
these schemes, one must not make any assumption on the global
state shared by Alice and Bob. That is, one must consider the most
general correlated state of~$N$ pairs of systems compatible with
the single-pair description. In our analysis, however, it is
assumed that Alice and Bob's state consists of~$N$ copies of the
same two-qubit state. In the prepare and measure picture, this
means that Eve interacts individually with the quantum states sent
to Bob; they are the so-called \emph{collective attacks}
\cite{KGR}. The recent results of \cite{KGR} suggest that Eve
gains no advantage by introducing correlations among the pairs of
systems shared by Alice and Bob. If this were proved correct, our
results would indeed provide a tight, general security proof for a
whole family of schemes and channels. Note also that while the
sufficient part of our security condition relies on the $N$ copies
hypothesis, the necessary part does not. It simply senses the
existence of an attack that can be applied to {\em any} protocol
equivalent to ours.

Finally, it is interesting to compare our results with previous
security proofs using two-way communication for QKD schemes
\cite{Chau}. When the attack described above is applied to the
six-state protocol, Eve prepares $N$ independent copies of a
two-qubit Werner state:
  \be
  \rho_{AB}= \lambda_1 \left[\Phi^+ \right]+\frac{1-\lambda_1}{3}\left( \left[\Phi^- \right]+
  \left[\Psi^+ \right]+\left[\Phi^- \right] \right) .
  \label{werner}
  \ee
Condition~(\ref{condition}) shows that a secure key extraction is
not possible with our protocol if the error rate is larger than
  \be
  \mbox{QBER}=2\frac{1-\lambda_1}{3} = 0.2764 .
  \ee
This is precisely the same value as obtained by Chau in
\cite{Chau}. Indeed, his protocol is equivalent to ours. The
attack we have presented proves that, unless another two-way
reconciliation technique is employed, this critical error rate
cannot be improved, i.e., is tight.

\section{acknowledgements}
%
%This work is supported by the Spanish MCyT, under  the ``Ram\'on y
%Cajal" grant, the 2002FI-00373,  2004FI-00068 and  BFM2002-02588 grants, the
%Generalitat de Catalunya (CIRIT project SGR-00185), the U.K. Engineering and Physical Sciences
%Research Council (IRC QIP), ...

This work is supported by the Spanish Ministry of Science and
Technology project BFM2002-02588, ``Ram\'on y Cajal", 2002FI-00373
and  2004FI-00068 grants, by CIRIT project SGR-00185, by the U.K.
Engineering and Physical Sciences Research Council (IRC QIP), and
by QUPRODIS working group EEC contract IST-2001-38877.


\begin{thebibliography}{99}
\bibitem{permutations} Any permutation in the Bell basis can be obtained by composing the following three transpositions:
  \bea
  &T(\Phi^+ \leftrightarrow \Phi^-)=&
  \left( \matrix{1&0 \cr 0&i} \right) \otimes \left( \matrix{1&0 \cr 0&i} \right)
  \nonumber \\
  &T(\Phi^- \leftrightarrow \Psi^+)=&
  \left( \matrix{\frac{1}{\sqrt{2}} & \frac{1}{\sqrt{2}}
  \cr \frac{1}{\sqrt{2}} & -\frac{1}{\sqrt{2}}} \right) \otimes
  \left( \matrix{\frac{1}{\sqrt{2}} & \frac{1}{\sqrt{2}}
  \cr \frac{1}{\sqrt{2}} & -\frac{1}{\sqrt{2}}} \right)
  \nonumber \\
  &T(\Psi^+ \leftrightarrow \Psi^-)=&
  \left( \matrix{1&0 \cr 0&-i} \right) \otimes \left( \matrix{1&0 \cr 0&i} \right)
  \nonumber
  \eea
\bibitem{mutinf} The quantum mutual information between systems $X$ and $Y$ is defined as
$ I(X:Y)=H(X)+H(Y)-H(X,Y) $,
where $H(X)=\mbox{tr}\! \left( \rho_X \log_2 \rho_X \right)$.
\end{thebibliography}
\end{document}